# MEMS Vapor Cells-based Rydberg-atom Electrometry Toward Miniaturization and High Sensitivity

Yintao Ma[a,b*], Beibei Sun[c], Pan Chen[a,b], Yao Chen[a,b*], Yanbin Wang[a,e], Ju Guo[a,e], Mingzhi Yu[a,b], Ping Yang[a,e], Qijing Lin[a,b,d], Libo Zhao[a,b,d*]

[a] State Key Laboratory for Manufacturing Systems Engineering, State Industry-Education Integration Center for Medical Innovations, International Joint Laboratory for Micro/Nano Manufacturing and Measurement Technologies, Shaanxi Innovation Center for Special Sensing and Testing Technology in Extreme Environments, Shaanxi Provincial University Engineering Research Center for Micro/Nano Acoustic Devices and Intelligent Systems, Xi'an Jiaotong University, Xi'an 710049, China

[b] School of Instrument Science and Technology, Xi'an Jiaotong University, Xi'an 710049, China

[c] East China Research Institute of Photo-Electron ICs., Bengbu, 233042, China

[d] Shandong Laboratory of Advanced Materials and Green Manufacturing at Yantai, Yantai 264000, China

[e] School of Mechanical Engineering, Xi'an Jiaotong University, Xi'an 710049, China

*corresponding authors (E-mails: yintaoma@xjtu.edu.cn, yaochen@xjtu.edu.cn and libozhao@xjtu.edu.cn)

## Abstract

Rydberg-atom electrometry, as an emerging cutting-edge technology, features high sensitivity, broad bandwidth, calibration-free operation, and beyond. However, until now the key atomic vapor cells used for confining electric field-sensitive Rydberg atoms nearly made with traditional glass-blown techniques, hindering the miniaturization, integration, and batch manufacturing. Here, we present the wafer-level MEMS atomic vapor cells with glass-silicon-glass sandwiched structure that are batch-manufactured for both frequency stability and electric field measurement. We use specially customized ultra-thick silicon wafers with a resistivity exceeding 10,000 $\Omega \cdot cm$, three orders of magnitude higher than that of typical silicon, and a thickness of 6 mm, providing a 4-fold improvement in optical interrogation length. With the as-developed MEMS atomic vapor cell, we configured a high-sensitivity Rydberg-atom electrometry with the minimal detectable microwave field to be 2.8 mV/cm. This combination of miniaturization and sensitivity represents a significant advance in the state-of-the-art field of Rydberg-atom electrometry, paving the way for chip-scale Rydberg-atom electrometry and potentially opening up new applications in a wider variety of fields.

## Introduction

The detection and sensing of microwave electric fields is of great significance in a variety of fields [1-3], including communications, military security, and astronomy. With the revolutionary development of quantum technology, particularly the advent of semiconductor tunable lasers enabling full advantage to be taken of resonance effects, the microwave electric field quantum precision measurement based on Rydberg atoms [4-7], regarded as a cutting-edge technology, has come into being. Rydberg-atom electrometry with exceptional sensitivity to external electric fields, a property attributed to its large polarizability ($\sim n^7$, where $n$ is the principal quantum number) and microwave transition dipole moment ($\sim n^2$), has demonstrated tremendous application potential in terms of precision [8], sensitivity [9], broadband tunability [10, 11], and subwavelength resolution spatial electric field imaging [12, 13]. Consequently, the Rydberg-atom electrometry are gradually replacing traditional metal dipole antennas, and it have attracted considerable attention and made leapfrog progress over the past decade or so.

The alkali-metal atomic vapor cells, acting as a hermetically sealed transparent container for confining Rydberg atoms, function as the core sensitive component of a Rydberg-atom electrometry. However, almost all atomic vapor cells currently available for Rydberg-atom electrometry are manufactured using traditional glass-blown techniques [14-17], which severely hinders the performance of this kind of sensor with respect to miniaturization, integration and scalability.

Following the general upward trend towards chip-scale integration and batch manufacturing, there is a growing endeavour to confine vapor atoms within well-defined geometries to achieve downscaled, and low-power light-vapor interactions. The quantum sensing technology empowered by Micro-Electro-Mechanical System (MEMS) is gradually turning this prospect into reality [18, 19]. The first and most essential phase is the microfabrication of chip-sized alkali-atom vapor cells using the state-of-the-art MEMS technology. Actually, micromachined alkali alkali-metal vapor cells have facilitated the implementation of miniaturized quantum devices [20-24], such as chip-scale atomic clocks, gyroscopes and magnetometers, significantly decreasing the size, weight, and power consumption of these quantum devices.

Nevertheless, two notable circumstances restrict the miniaturization of the Rydberg atomic system to a considerably lower integration degree than other alkali-metal atom-based quantum sensors. The extremely sensitive Rydberg state demands an ultra-high vacuum vapor cells for preventing spin quantum state decoherence. Additionally, strict restrictions are also imposed on the materials used to fabricate the vapor cells for the purpose of maintaining the fidelity of the microwave fields, thereby preventing distortion phenomena, such as absorption and scattering. Despite existing challenges, there have been preliminary and sporadic attempts to incorporate wafer-level MEMS vapor cells into Rydberg-atom electrometry in recent two years [25, 26]. The typical glass-silicon-glass triple-layer stacked structure vapor cells were successfully used for measurement of microwave electric field. However, the limited optical interrogation length and low resistivity defined by silicon wafer result in low sensitivity and accuracy. The all-glass wafer-level vapor cells have also been developed specifically for millimeter-wave electric fields measurement. Although the full-dielectric vapor cells offer advantages in terms of microwave field loss and shielding aspects, there are still several undesirable drawbacks compared to anode-bonded vapor cells, including mandatory cleaning and activation procedures (a mixture of concentrated sulfuric acid and hydrogen peroxide for 1 hour), high bonding temperature (~500 °C), low processing efficiency (~20 hours), and strict processing treatment intervals (within 30 minutes).

In regard to the aforementioned problems, we successfully develop extended optical access, ultra-high vacuum, and wafer-level MEMS alkali-metal atomic vapor cells for Rydberg-atom electrometry leveraging hybrid MEMS manufacturing processes. The tailor-made silicon wafer with a thickness of 6 mm and a resistivity of 10,000 Ω·cm and borosilicate glass serve as microwave field-friendly materials for the microfabrication of MEMS vapor cells. Even more noteworthy is that we replace all traditional glass-blown vapor cells with the as-developed MEMS vapor cells, including frequency stabilization and electric field sensing. We extensively characterize Doppler saturation absorption and electromagnetic induction transparency (EIT) utilizing two types of micro-fabricated vapor cells with different geometries, and implement a highly sensitive Rydberg-atom electrometry via resonance-driven $|48\,D_{5/2}\rangle$ to $|46\,F_{7/2}\rangle$ Rydberg state transitions.

## Materials and methods

### Microfabrication of MEMS atomic vapor cells

In contrast to the MEMS atomic vapor cells typically utilized in atomic magnetometers or clock, here we have incorporated improvements both in terms of material selection and manufacturing process for those used in Rydberg-atom electrometry. The microfabricated vapor cells are constructed by twice anodic bonding processes, thereby providing the hermetically sealed glass-silicon-glass sandwich structure. The thickness of the intermediate silicon core is 6 mm, which defines the optical length of the interrogated atom. The detailed fabrication procedure is indicated in Fig. 1.

Specially customized 6-mm-thick double-polished silicon wafers with a resistivity of 10,000 Ω·cm and 500-μm-thick borosilicate glass (BF33) are provided as raw materials for the MEMS atomic vapor cells. That kind of silicon wafers with ultra-high resistivity features negligible magnetic noise and may provide less disturbance to radio frequency (RF) fields to be measured.

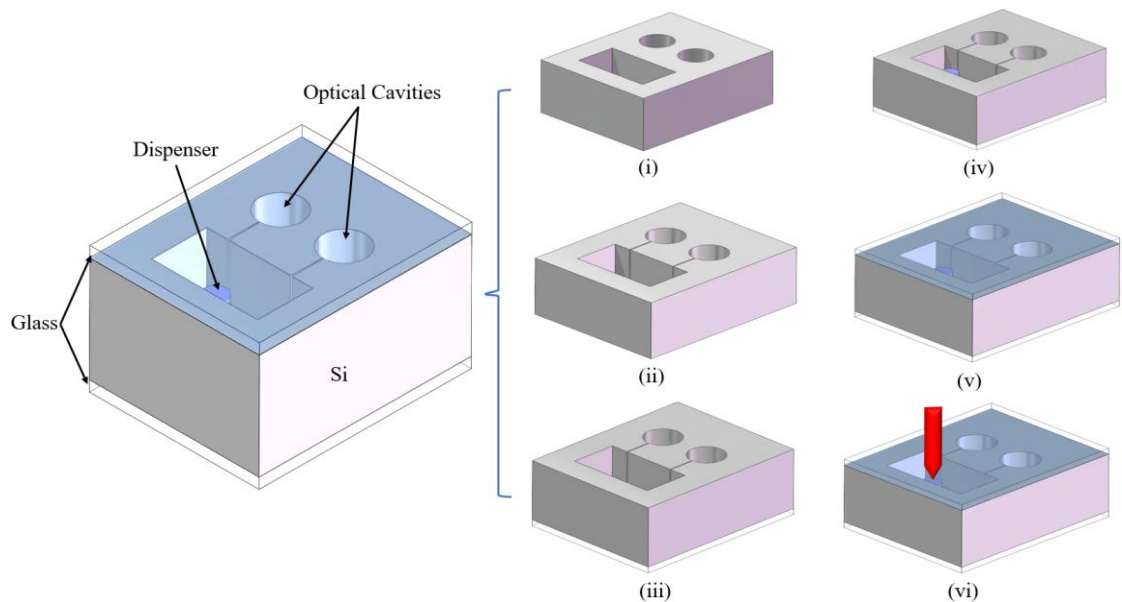

Fig. 1. Microfabrication procedure of wafer-scale atomic vapor cells.

(i) An array of silicon through-holes is drilled into the silicon wafer using mechanical processing method. A protective layer is deposited on the silicon surface prior to processing as much as possible to minimize chipping effect around the silicon holes. Each set of two or three holes works as a group. For double-hole design, one acts as an optical cavity and the other as a reservoir cavity, used for saturation absorption experiments. While for three-hole arrangement, two of them are optical cavities and one is a reservoir cavity, designed for optical differential detection of Rydberg-atom electrometry.

(ii) The optical cavities and reservoir cavities are spatially connected via microchannels, which are created using femtosecond laser machining technology. The cross-sectional dimensions of the microchannel are 800 μm × 200 μm, allowing alkali-metal atoms to diffuse into the optical cavities while preventing impurities generated by the Cs dispenser ($Cs_2CrO_4$/Zr/Al in pill) from contaminating these optical cavities.

(iii-iv) Next, place the ready-processed silicon wafer and glass wafers to be bonded in piranha solution (7:3 of concentrated sulphuric acid and hydrogen peroxide) with ultrasonic cleaning treatment for 10 min to thoroughly purify the surface. The first glass-silicon bonding is carried out under atmospheric conditions of 300 °C and 1 kV to form the semi-enclosed chambers. Following that, the commercially available Cs dispenser, a thermally activated compound that produces Cs atoms, is distributed into each reservoir cavity individually.

(v) The semi-enclosed chambers loaded with the Cs dispenser and the top BF33 glass wafer are placed in the bonding system heated to 300 °C for 2 hours for degassing treatment, while continuously evacuating the system using a molecular pump, reaching a vacuum level of $\sim 10^{-5}$ Pa before sealing the vapor cells. Then, the second bonding is accomplished by applying a voltage of 1 kV for 20 minutes, thereby forming the three-layer stacked wafer-level MEMS atomic vapor cells, as shown in Fig. 2 (a).

(vi) The single cell chip is subsequently obtained by dicing the wafer-level vapor cells. Finally, the Cs vapor atoms are chased into the optical cavity after being thermally activated of Cs dispenser by a 1550 nm infrared laser beam. The functionally complete double-chamber and triple-chamber atomic vapor cells are shown in Fig. 2 (b) and 2 (c), respectively.

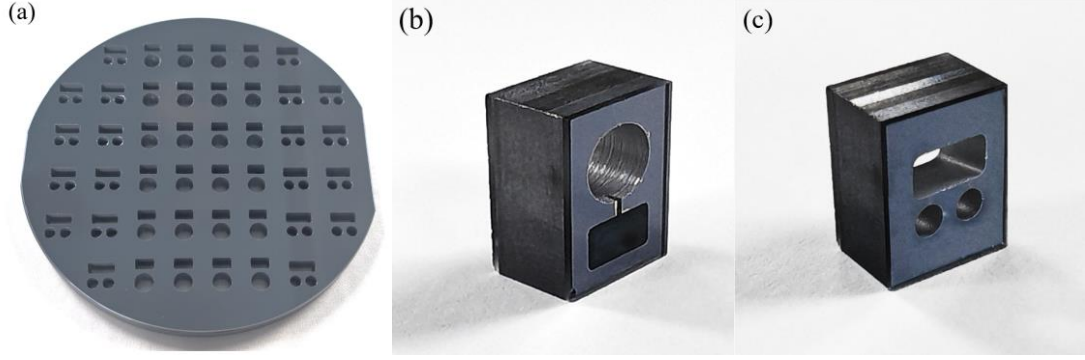

Fig. 2. Microfabricated MEMS atomic vapor cells. (a) wafer-level vapor cells. (b) double-chamber vapor cell chip for laser frequency stabilization. (c) triple-chamber vapor cell chip for RF electric field measurement.

**Experimental setup**

The schematic of the Rydberg-atom electrometry is depicted in Fig. 3, primarily consisting of as-fabricated MEMS vapor cells, laser systems (probing laser and coupling laser), a frequency-stabilization system, a differential detection and microwave system.

We use the MEMS vapor cells not only to confine Rydberg atoms for electric field measurement, but also to generate saturated absorption spectroscopy (SAS) for laser frequency stabilization. Specifically, for Rydberg-atom electrometry, the three-chamber MEMS vapor cell features two identical optical cavities with the internal dimension of $\Phi\ 3 \times 6$ mm$^3$ shared by the same experimental conditions, facilitating better differential detection; The MEMS vapor cell used for saturated absorption frequency stabilization has only one optical cavity with the internal dimension of $\Phi\ 6 \times 6$ mm$^3$.

Using a two-photon excitation scheme, the Cs atoms are excited from the ground state to the Rydberg state, i.e., the 852 nm probing beam drives the transition of $|6\ S_{1/2}\rangle \rightarrow |6\ P_{3/2}\rangle$, while the 510 nm coupling beam drives the transition of $|6\ P_{3/2}\rangle \rightarrow |48\ D_{5/2}\rangle$, thereby a forming ladder-type three-level configuration. Following half-wave plate (HWP) and polarizing beam splitter (PBS), the 852 nm laser beam is separated into two portions, one of which is then further equally divided into two beams, concurrently transiting through two optical cavities of the three-chamber MEMS vapor cell for interrogating Rydberg atoms, while the other minor portion is directed into the frequency-stabilization system, locking the probing frequency at $|6\ S_{1/2}\ (F=4)\rangle \rightarrow |6\ P_{3/2}\ (F'=5)\rangle$ (See Section 4.1 for the details). After being guided by a dichroic mirror (DM), the 510 nm laser beam coincides with the probing beam and propagates in the opposite direction through the MEMS vapor cell. The beam sizes of the probing beam and coupling beam are 500 μm and 600 μm, respectively. Then, two beams of light transmitted through the MEMS vapor cell hit the balance detectors (BD), enabling differential detection for suppressing common-mode noise, such as light fluctuations, etc. In the experiment, the narrowband pass filters are also mounted in front of the BD to further reduce the adverse effects of stray light field from the surrounding environment.

The sensitive MEMS vapor cell is immersed in a microwave field, the frequency of which is tunable to 29.75 GHz, resulting in the coupling transition of adjacent Rydberg states $|48 D_{5/2}\rangle \rightarrow |46 F_{7/2}\rangle$. This transition manifests itself spectroscopically as an Autler-Townes (AT) splitting. Finally, the EIT and AT splitting signal is displayed and collected by an oscilloscope.

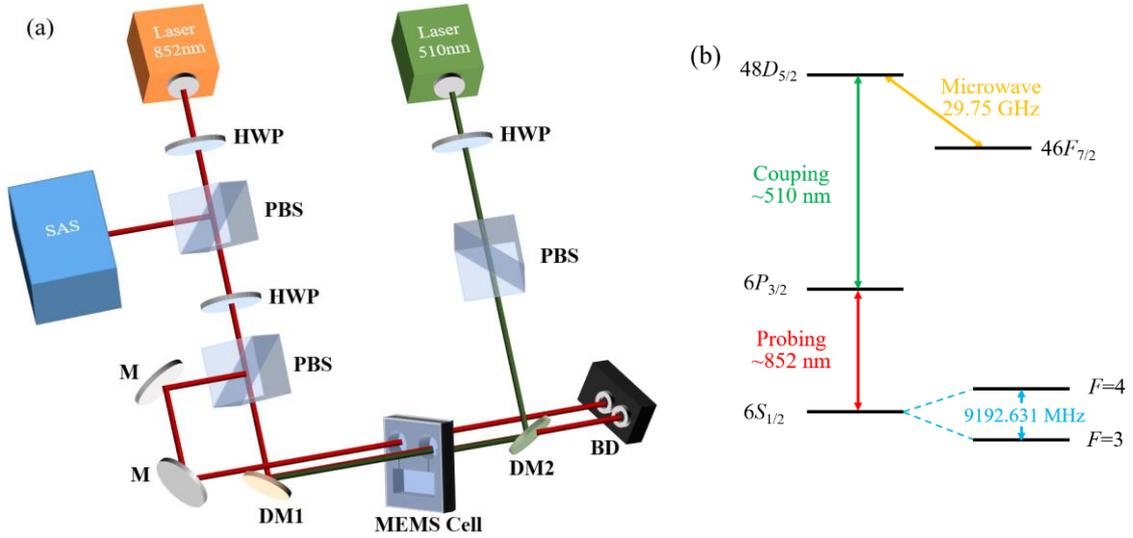

Fig. 3. Schematic of the Rydberg-atom electrometry. (a) Experimental setup. Three-chamber MEMS atomic vapor cell with two optical cavities serves as the core sensitive component for differential detection of Rydberg-atom electrometry. HWP: half-wave plate; PBS: polarizing beam splitter; M: mirror; DM: dichroic mirror; BD: balance detectors; SAS: saturated absorption spectroscopy. (b) Energy-level diagram of Rydberg atoms.

## Experimental results
### Saturated absorption spectroscopy (SAS)

Maintaining the stabilized frequency of the probing laser is the most critical aspect for capturing EIT signal. The narrower the frequency detuning of the probing laser, the more pronounced the EIT signal becomes [27]. In this experiment, the SAS was adopted for frequency stabilization of the probing laser. Meanwhile, we removed the glass-blown vapor cell traditionally used in saturation absorption frequency-stabilized systems and substituted it with the newly developed dual-chamber MEMS vapor cell to verify the performance. The detailed experimental set-up and procedure can be found in [28, 29]. Fig. 4 clearly shows the measured SAS. The six hyperfine transitions corresponding to $|6\,S_{1/2}\,(F=4)\rangle \rightarrow |6\,P_{3/2}\,(F'=3)\rangle$, $|6\,S_{1/2}\,(F=4)\rangle \rightarrow |6\,P_{3/2}\,(F'=3\text{-}4)\rangle$, $|6\,S_{1/2}\,(F=4)\rangle \rightarrow |6\,P_{3/2}\,(F'=4)\rangle$, $|6\,S_{1/2}\,(F=4)\rangle \rightarrow |6\,P_{3/2}\,(F'=3\text{-}5)\rangle$, $|6\,S_{1/2}\,(F=4)\rangle \rightarrow |6\,P_{3/2}\,(F'=4\text{-}5)\rangle$, and $|6\,S_{1/2}\,(F=4)\rangle \rightarrow |6\,P_{3/2}\,(F'=5)\rangle$, respectively, are referred to as three intrinsic absorption peaks and three cross-absorption peaks. The successful resolution of individual hyperfine transition peaks within the spectra demonstrates the vapor cell's suitability. Here, the laser frequency is actively locked to the intrinsic absorption peak at $|6\,S_{1/2}\,(F=4)\rangle \rightarrow |6\,P_{3/2}\,(F'=5)\rangle$, keeping the frequency stability better than 5 MHz.

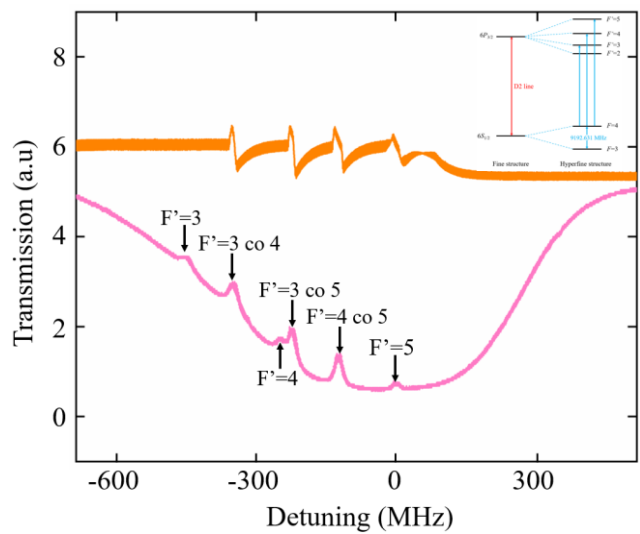

Fig. 4. SAS of Cs atoms D2 line in MEMS atomic vapor cell. The inset figure shows the energy level structure of Cs atoms.

**EIT signal and differential detection**

We would like to begin by providing a brief description of how EIT signals are detected before conducting RF-field measurement experiments. The optical differential detection scheme with a three-chamber MEMS vapor cell, i.e., subtracting the reference light beam travelling through one of optical cavity from the functionalized light beam transmitting through another optical chamber, is used to suppress common-mode noise.

The typical EIT signals when scanning coupling light frequency are shown in Fig. 5, from which it can be very clearly observed that the signal-to-noise ratio is greatly improved by a factor of approximately 20 for $|48\,D_{5/2}\rangle$ Rydberg state.

Following that, we measured the EIT spectrum operating in differential detection arrangement and investigated the influence of experimental parameters (probing and coupling power) on the EIT spectrum, as presented in Fig. 6 and Fig. 7. Fig. 6 (a) illustrates several representative EIT spectral lines subjected to varied probing light powers ranging from 20 to 300 μW. These experimental data can be overlaid with a double-peak Lorentz function, from which we can extract the linewidth and amplitude of the EIT spectrum,

$$\sum_{i=1}^{2} S_{EIT}(f_c) = \frac{A_i \Gamma_i}{(f_c - f_{0i})^2 + \Gamma_i^2} + B_i \tag{1}$$

where $\Gamma$ is EIT linewidth, $f$ is resonant frequency, $A$ and $B$ are the fitting coefficients, respectively.

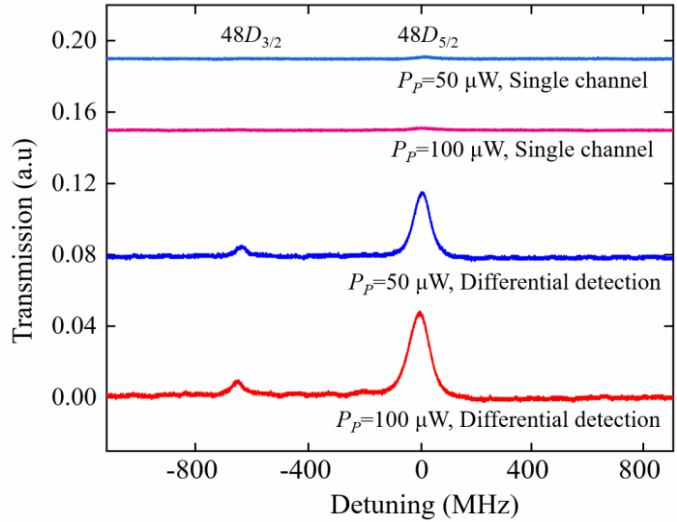

Fig. 5. Experimentally measured typical EIT signals of differential detection and single-channel configurations under different probing power.

According to the fitting results of Eq. (1), the dependencies between the EIT spectral amplitude, linewidth and probing power are shown in Fig. 6 (b) and Fig. 6 (c), respectively. In Fig. 6 (b), the EIT amplitude gradually grows with the probing power and approaches saturation, indicating that the excitation of the probing light triggers the Cs atoms in the Rydberg state to increase, then limited by the effective number of atoms at room temperature. Fig. 6 (c) shows the monotonic rise in the relationship between EIT linewidth and probing power. The dependence of measured linewidth on probing intensity is due to optical broadening effect. Using the theoretical model $\Gamma(P_p) = C\sqrt{P_p} + \Gamma_0$, (where $P_p$ is the probing power, $\Gamma_0$ is the intrinsic EIT linewidth, $C$ is the fitting coefficient), the intrinsic linewidth $\Gamma_0$ extrapolated to zero optical power is approximately 10.1 MHz. It should also be clarified that even though there are two Rydberg states, $|48\ D_{5/2}\rangle$ and $|48\ D_{3/2}\rangle$ within the scanning frequency range, in the displayed spectrum, here we concentrate on $|48\ D_{5/2}\rangle$, the one that has a more prominent signal intensity, in Fig. 6 (b) and Fig. 6 (c).

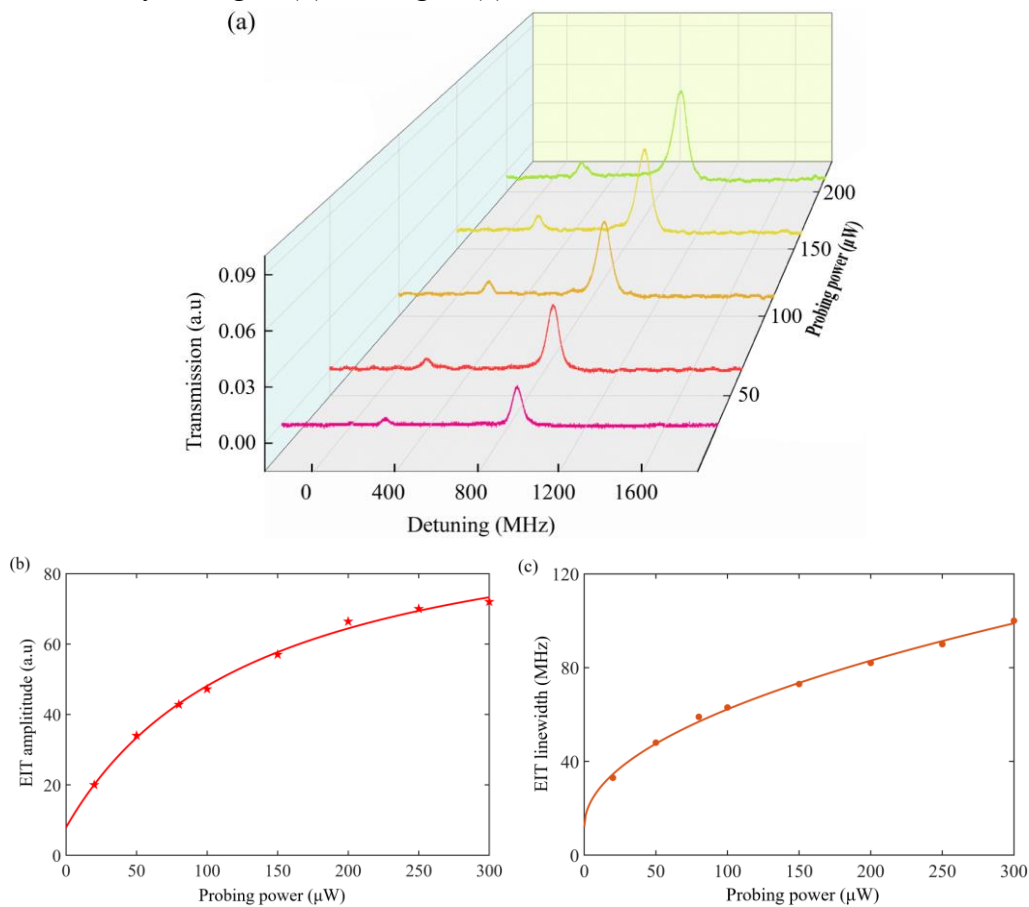

Fig. 6. (a) EIT spectrum under different probing powers at a fixed coupling power of 50 mW. (b) EIT amplitude as a function of probing powers. (c) EIT linewidth versus probing powers.

**Rydberg-atom electrometry**

Next, we configured a Rydberg-atom electrometry for measuring the electric field using the developed MEMS atomic vapor cell. The microwave field with a frequency of 29.75 GHz coupling the $|48\ D_{5/2}\rangle$ Rydberg state to the $|46\ F_{7/2}\rangle$ state acts on the vapor cell through a horn antenna, resulting in a cascaded four-level atomic system. From a spectroscopic point of view, the EIT spectrum would split into two peaks, generally referred to as EIT-AT splitting effect. The Rabi frequency $\Omega_{MW}$ corresponding to the microwave electric field interacting with adjacent Rydberg energy states is expressed as,

$$\Omega_{MW} = \frac{\mu|E|}{\hbar} \quad (2)$$

The separation of the EIT-AT splits $\Delta f$ is equivalent to the Rabi frequency $\Omega_{MW}$, i.e., $\Omega_{MW}=2\pi\Delta f$. Therefore, the separation of the splits $\Delta f$ is directly proportional to the microwave amplitude $|E|$ and can be formulated as the following equation [30-32],

$$|E| = \frac{\hbar}{\mu}\Omega_{MW} = \frac{2\pi\hbar}{\mu}\Delta f \quad (3)$$

where $\mu$ is atomic dipole moment, and $\hbar$ is the reduced Planck constant.

The experimental observations of the EIT-AT splitting dressed by microwave fields are shown in Fig. 7. Fig. 7 (a) highlights the comprehensive results of the spectral pattern subjected to the applied microwave fields, including EIT-AT splitting and transmission amplitude. Two apparent phenomena, namely, the higher microwave power leads to the more pronounced spacing and lower transmission signals, can be discovered. In order to further evaluate the relationship between EIT-AT splitting and the microwave fields, we extracted experimental data from Fig. 7 (a) and carried out linear fitting using Eq. (3), as presented in Fig. 7 (b), from which a linear fitting slope of 46.3 MHz/$\sqrt{mW}$ can be acquired. Furthermore, the minimum detectable microwave field in the MEMS vapor cell-based Rydberg-atom electrometry can be as low as 2.8 mV/cm. Fig. 7 (c) demonstrates how the EIT-AT signal amplitude is influenced by microwave fields. The experimental curve indicates a general trend of attenuation with changes in microwave power. When microwave fields are at lower level, the EIT-AT amplitude remains relatively stable and changes slowly, revealing that the changes in microwave power have a small impact on the EIT-AT effect over this range; With microwave power increasing even more, the EIT-AT amplitude decreases speedily, suggesting that microwave power has been enhanced to a certain state, significantly interfering with the initial EIT-AT process, resulting in a notable reduction in the signal amplitude.

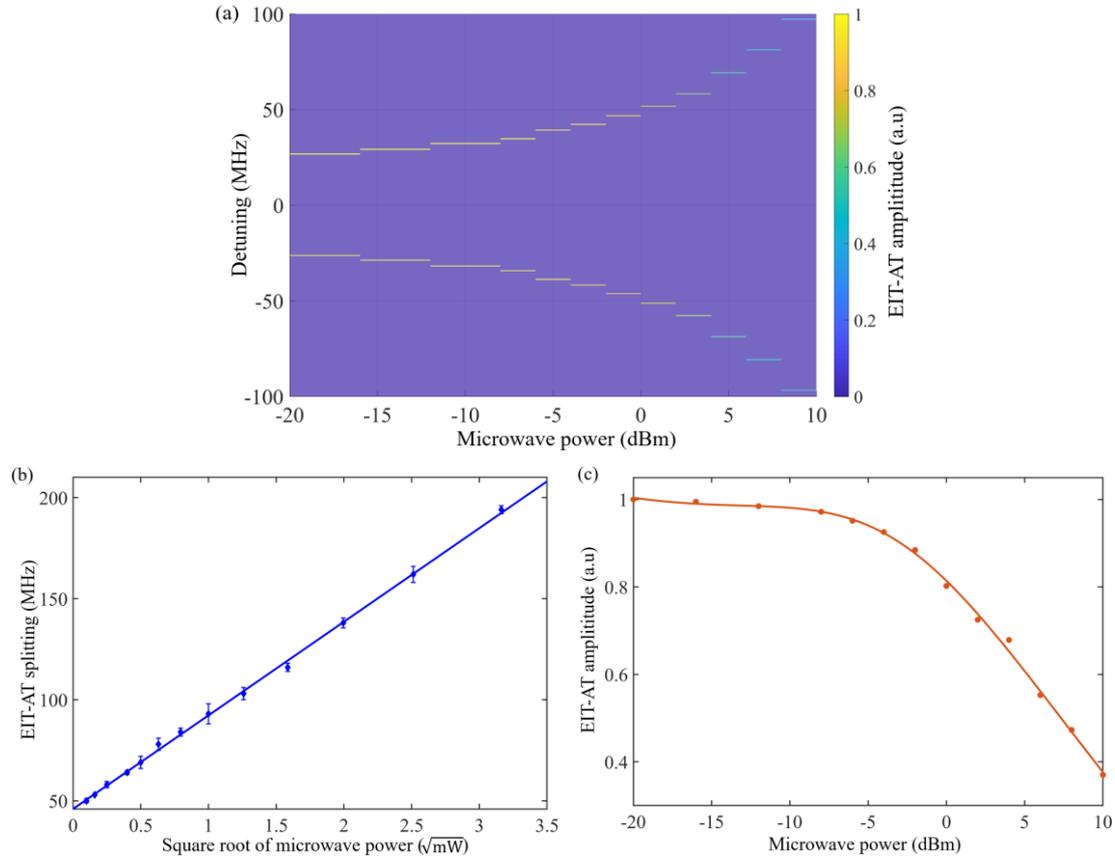

Fig. 7. Four-level EIT spectrum driven by microwave fields. (a) Visualizations of EIT spectrum as a function of the microwave power and coupling detuning. (b) Linear dependence of the EIT-AT separation on the square root of the microwave power, from which a slope of 46.3 MHz/$\sqrt{mW}$ can be ascertained. (c) Normalized EIT-AT amplitude response curve.

## Discussion and Conclusion

This study successfully demonstrates a significant advancement in Rydberg-atom electrometry through the development and application of wafer-level MEMS atomic vapor cells. By replacing all traditional glass-blown vapor cells previously used in the Rydberg-atom electrometry with glass-silicon-glass sandwich structure micro-fabricated MEMS vapor cells, we simultaneously achieve miniaturization, batch manufacturability, and enhanced sensitivity—key steps for transitioning quantum electrometry toward chip-scale integration.

The MEMS vapor cell serves two roles: (i) as the core component for microwave electrometry and (ii) for laser frequency stabilization via SAS, eliminating the need for traditional glass cells and streamlining system architecture. The use of ultra-high resistivity, RF field-friendly silicon (10,000 Ω·cm, 6 mm thick) minimizes RF field distortion and magnetic noise while quadrupling the optical interrogation length. This addresses the critical trade-off between Rydberg-atom electrometry in term of sensitivity and miniaturization.

However, several limitations remain. First, stray electric fields from Cs dispenser ($Cs_2CrO_4$/Zr/Al), despite mitigation via microchannel separation, still may cause observable shifts in EIT resonance peaks. This problem may be alleviated by further optimization of the microchannel geometry and alkali-metal filling methods. Second, relatively high bonding temperature (~300°C) limits material choices. Low-temperature or even room-temperature bonding processes should be further explored. In addition, in-depth research on anti-relaxation coating inside the vapor cells compatible with MEMS fabrication processes may bring some unexpected benefits for the Rydberg-atom electrometry.

In conclusion, this paper establishes a viable pathway for miniaturized, high-sensitivity Rydberg-atom electrometry using MEMS technology. By harmonizing ultra-high resistivity materials, extended optical access, wafer-scale fabrication, and differential detection, we overcome the long-standing barriers of miniaturization while maintaining the high sensitivity. The demonstrated 2.8 mV/cm sensitivity and batch manufacturability pave the way for the promising chip-scale quantum electrometry. Future efforts will focus on integrating auxiliary components onto the same chip via hybrid integration with photonics technology to realize fully monolithic quantum RF-field sensors.

## Disclosures
The authors declare no conflicts of interest.

## Data availability statement
The experimental data underlying the results presented in this paper are not publicly available at this time but may be obtained from the corresponding authors upon reasonable request.

Yintao Ma and Beibei Sun contributed equally to this work.

## Acknowledgments
This work was supported in part by the National Key Research & Development (R&D)

Program of China (Grant No. 2024YFB3212500), National Natural Science Foundation of China (Grant No. 52435010), and the Shaanxi Provincial Science and Technology Development Program (Grant Nos. 2023-LL-QY-35, 2024RS-CXTD-19).